# Secure Numerical and Logical Multi Party Operations

Johannes Schneider[a], Bin Lu[b]

[a]*University of Liechtenstein, Liechtenstein*
[b]*Ecole Polytechnique Federale de Lausanne, Switzerland*

**Abstract**

We derive algorithms for efficient secure numerical and logical operations in the semi-honest model ensuring statistical or perfect security for secure multi-party computation (MPC). To derive our algorithms for trigonometric functions, we use basic mathematical laws in combination with properties of the additive encryption scheme, ie. linear secret sharing, in a novel way for the JOS scheme [23]. For division and logarithm, we use a new approach to compute a Taylor series at a fixed point for all numbers. Our empirical evaluation yields speed-ups for local computation of more than a factor of 100 for some operations compared to the state-of-the-art.

## 1. Introduction

Consider the following tasks: i) Identify people on a picture without looking at it; ii) Outsource computations giving away encrypted data but keeping keys private. Both tasks come with the challenge that there is no access to the non-encrypted data. It seems impossible to work on encrypted data only. Surprisingly, computing on encrypted data is indeed doable. A rather mature technique is secure multi-party computation (MPC) relying on non-collusion of a network of parties. To this date, MPC suffers heavily from its performance overhead. Whereas a lot of emphasis has been put on optimizing the computation of Boolean circuits, only limited effort has been made to secure numerical operations efficiently. For example, prior work did not deal with trigonometric functions such as sine or cosine needed in many applications, such as signal processing in an industrial context. In fact, aside from basic operations (such as addition and multiplication) no complex mathematical operation can be carried out efficiently and accurately. Prior work uses either (slow) iterative schemes or approximations of the function to compute. We address this gap using a recent scheme called JOS [23] that explicitly separates between keys and encrypted values. It supports various variants of linear secret sharing, eg. additive blinding with and without modulo as well as using XOR. The distinction between keys and encrypted values together with the simple encryption schemes lend itself well to make use of basic mathematical equations that relate the ciphertext, the plaintext and the key. In essence, to compute some functions we can use the same implementation (plus a few additional operations) used for plaintexts on the ciphertexts or keys as we show for trigonometric functions. This makes it possible to benefit from the long history of optimizations of implementations and algorithms for non-encrypted data. For illustration, our empirical evaluation yields that the amount of local computation per party to compute a sine function is only about a factor 2 more than for computation on non-encrypted data. At times, we also employ the idea of using multiple encryptions of the same plaintext to derive a system of equations to leverage operations on non-encrypted data. This helps to deal with a reduced key space caused by the inability to evaluate certain functions designed for non-encrypted data on arbitrary keys. Additionally, we discuss a method for computing Taylor series based on scaling the secret value. The scaling makes it possible to develop the series at a fixed number (for the entire value range of a secret). This approach yields fast conversion for a broad range of functions as we demonstrate for division and logarithm.

For logical operations, the key ingredient is an efficient comparison protocol for equality (with zero) and for checking if a value is less than zero. This is done by using algorithms for conversions between encryption schemes and using large Fan-In gates. Our ideas might prove valuable in other settings or using other schemes aside from JOS well.

## 1.1. Contributions

- Presenting the first algorithms for efficient computation of trigonometric functions, ie. sine, cosine and tangent. They provide statistical security using only five rounds, local computation proportional to computation without encryption and communication of $O(k)$ bits where $k$ is the security parameter. They improve on series-based techniques [14] by more than a factor of 10 in local computation and communication.

- Stating an algorithm for calculating Taylor series efficiently for a wide range of functions, demonstrated for division and logarithm. More concretely, we improve the round complexity of the state-of-the-art [1, 7] for division and logarithm for computation on 32-bit floats and 64-bit double values by more than 10 rounds.

- Presenting an algorithm for division of a confidential number by a public divisor requiring only one round without the need to perform comparisons, which take significantly more than one round [6, 27].

- Introducing a number of efficient operations using the JOS scheme [23] for comparison(equality, less than), conversion between different forms of encryptions and large fan-in gates that achieve comparable or better performance to prior work using different schemes. They require a constant number or almost constant rounds, ie. $O(\log^* l)$ and $O(\log_b l)$, where $l$ is the number of bits of the encrypted value and $b$ a parameter. In terms of local computation our equality protocol is more than a factor 100 faster than the state-of-the-art.

*1.2. Outline*

After some preliminaries (Section 1.3) focusing on summarizing the JOS scheme and presenting some notation and conventions, we introduce algorithms for three areas: conversions between encryption schemes (Section 2), logical operations (Section 3) and numerical operations (Section 4). There are some interdependencies between algorithms from different sections, eg. some conversion algorithms between encryption schemes are used by some algorithms for logical operations. Finally, we give a short empirical evaluation (Section 5) and we discuss related work (Section 6).

*1.3. Preliminaries and Notation*

We briefly recapitulate notation and concepts from the JOS scheme [23]. For a secret value $a \in [0, 2^l - 1]$ with $l$ bits and a key $K$ with $b \geq l$ bits we consider three kinds of linear encryptions:

- $ENC_K(a) = a + K$ : (Purely) additive encryption
- $ENC_K(a) = (a + K) \mod 2^l$: Additive (modulo) encryption
- $ENC_K(a) = a \oplus K$: XOR encryption

Given an encryption $ENC_K(a)$ we denote the effective maximum number of bits by the key or the encrypted value by $l_E$. For $ENC_K(a) := (a + K) \mod 2^l$, we have $l_E = l$. For additive encryption of a key $K \in [0, 2^b - 1]$ with $b$ bits, we have $l_E = b + 1$ for $a + K \geq 2^b$ and $l_E = b$, otherwise. Denote by subscript $i$ the $i^{th}$ bit of a number in little Endian notation, eg. for $a = 10$: $a_0 = 0$ and $a_1 = 1$. In particular, $E_i, a_i$ and $K_i$ denote the $i^{th}$ bit of $ENC_K(a), a$ and $K$. The JOS scheme [23] uses three parties, namely: a key holder (KH), an encrypted value holder (EVH) and a helper (HE). The KH holds keys only (most of the time), the EVH keeps encrypted values (most of the time) and the helper can have either of them, but it is not allowed to have an encrypted value and the matching key. For additive encryption $a + K$, we define the carry bit $c_i$ to be the "carry over" bit that is added to $a_{i+1}$ aside from $k_{i+1}$ during encryption. Thus, by definition $c_0 := 0$ and for $i > 0$ we have $c_i := 1$ iff $c_{i-1} + a_i + k_i > 1$, otherwise $c_i := 0$. Frequently, we encode 'TRUE' as one and 'FALSE' as zero. Our algorithms process as inputs encrypted values from the EVH and keys from the KH. They ensure the encrypted value of the result is held by the EVH and its key by the KH. For inputs and outputs we write an encrypted value and key as pair $(ENC_K(a), K)$. Thus, we write for a function $f$ operating on an encryption of value $a$ returning an encrypted value $rE$ and key $rK$ the following $(rE, rK) := f(ENC_K(a), K)$. In particular, we use the multiplication protocol $MUL$ [23] and the protocol for bitwise AND, ie. $AND$. Both typically take two confidential numbers as input. For $AND$ we sometimes use larger Fan-Ins as described in [23]. We also assume a protocol for computing the power $a^i$ for a non-confidential integer $i > 0$, ie. power $POW((ENC_K(a), K), i)$. It can be implemented using the multiplication protocol $MUL$ using $O(\log i)$ multiplications. To reduce the number of bits needed we also use scaled power computation $SCALEDPOW((ENC_K(a), K), i, s)$, ie. for a non-confidential scaling factor $s$ we compute $a^i/s^{i-1}$. We enumerate keys either by using primes, eg. $K', K'', K'''$ or using numbers $K^0, K^1, K^2$. The terms $E'$, $E''$ and $E'''$ denote that an encrypted value with $K'$, $K''$ and, respectively, $K'''$.

## 2. Conversions between Encryptions

We show how to convert between all three encryption schemes, ie. XOR, additive with and without modulo.

### 2.1. Additive ↔ XOR Encryption

Algorithm AddToXOR computes an XOR encryption of a secret from an additive encryption(with or without modulo). It uses the carry bits algorithm in Section 3.5[1] and it exploits the definition of the carry bit $c_i$ to get XOR encryptions of the bits.

---
**Algorithm 1** AddToXOR(encrypted value $ENC_K(a)$, key $K$)
---
1: $(ENC_{K_i''}(c_i), K_i'') := CarryBits(ENC_K(a), K)$ with $i \in [0, l-1]$
2: $K_i' := K_i \oplus K_i''$ {by KH}
3: $ENC_{K_i'}(a_i) := e_i \oplus ENC_{K_i''}(c_i)$ {by EVH}
4: return $(ENC_{K_i'}(a_i), K_i')$
---

**Theorem 1.** *Algorithm 1 converts correctly and securely from additive to XOR encryption.*

*Proof.* For bit $e_i$ of $ENC_K(a) = a + K$ (with or without $\mod 2^l$) we have using the definition of the carry bit:

$$e_i = (a_i + k_i + c_i) \mod 2 = a_i \oplus k_i \oplus c_i \tag{1}$$

This can also easily be verified by listing all 8 options for each bit $a_i, k_i$ and $c_i \in \{0, 1\}$. In Algorithm AddToXOR we compute $e_i' := e_i \oplus e_i''$. We show that this definition ofo $e_i'$ is equivalent to claimed return value, ie. an XOR encryption

---
[1] Due to the dependencies among algorithms, it is not possible to avoid referencing later sections of the paper without removing a consequent structuring of the paper into three main sections (encryption conversions, logical and numerical operations).

of $a_i$ with key $k'_i$, ie. $a_i \oplus k'_i$ with $k'_i := k_i \oplus k''_i$. We have:

$a_i \oplus k'_i$
$= \quad a_i \oplus k_i \oplus k''_i \oplus 0 \quad$ (by definition of $k'_i$ and since $x \oplus 0 = x$)
$= \quad a_i \oplus k_i \oplus k''_i \oplus (c_i \oplus c_i) \quad$ (since $x \oplus x =$)
$= \quad (a_i \oplus k_i \oplus c_i) \oplus k''_i \oplus c_i \quad$ (due to commutativity of $\oplus$)
$= \quad ((a_i + k_i + c_i) \mod 2) \oplus e''_i \quad$ (by definition of $e''_i$ and Equation 1)
$= \quad e_i \oplus e''_i \quad$ (by Theorem 8)
$=: \quad e'_i$

Security follows from the security of Algorithm CarryBits (Theorem 8) and the fact that Algorithm AddToXOR does not cause any sharing of information between the KH, EVH and HE. □

Algorithm XORtoADD performing the opposite conversion is more involved. Both the KH and the helper make decisions depending on a key bit of the additive encryption. The helper assists in computing the encrypted bits of the result. It needs input that depends on the encrypted values as well as keys. Since it is not possible that the helper obtains a matching key and for an encrypted value, the EVH 'double' encrypts bits before sending them to the helper. The keys used for doubling encryption are not given to the HE but only to the KH. Since we compute each bit of the additively encrypted bits separately, we must combine them into one encrypted value. Clearly, for arithmetic rings this requires that all of them stem from the same arithmetic ring, eg. to add numbers $a + K \mod 2^l$ and $b + K' \mod 2^{l'}$, we should have that $l = l'$. Otherwise it is not clear, which modulo to take (e.g. $\mod 2^l$ or $\mod 2^{l'}$) for the added keys $K + K'$ and the added encrypted values. We address this problem by encrypting bit $i$ using a key with $2^{l-i}$ bits, and scaling by $2^i$ transforming all bits to the same arithmetic ring ($\mod 2^l$) before computing their sum.

The helper uses the following observations: If the key bit $K_i$ is zero then the plaintext equals the ciphertext, ie. $e_i = a_i$. If the key bit is one then $a_i = \neg e_i = 1 - e_i$. The key step is the last equation: We can express the Boolean negation operation in terms of an arithmetic operation. This is important, since the helper cannot directly compute $1 - e_i$ because it obtains $e_i$ additively encrypted, ie. $ENC_{K''_i}(e_i)$. However, it can compute $ENC_{K''^{-1}_i}(1 - e_i) = 1 - ENC_{K''_i}(e_i) + 2^{l-i} \pmod{2^{l-i}}$. Since the KH also knows $K_i$, it knows how bit $i$ got encrypted, ie. whether to use $K''^{-1}_i$ or $K''_i$. Note, that the helper cannot disclose any information to the EVH that depends on the key. Thus, it must encrypt its findings (using a key $K'''_i$) before returning results to the EVH.

**Theorem 2.** *Algorithm 2 converts correctly and securely from XOR encryption to additive encryption.*

*Proof.* Security follows since none of the three parties exchange an encrypted value or a key that allows any of them to decrypt a value: The EVH gets $ENC_{Kf(i)}(a_i)$ with $Kf(i) := K''_i + K'''_i \pmod{2^{l-i}}$ but it does not get

**Algorithm 2** XORToAdd(encrypted value $ENC_K(a)$, key $K$)

1: Choose keys $K_i'' \in [0, 2^{l-i}]$ {by EVH}
2: $ENC_{K_i''}(ENC_{K_i}(a_i)) = (a_i \oplus K_i) + K_i''$ $(mod\ 2^{l-i})$ {by EVH}
3: EVH sends $K_i''$ to KH and $ENC_{K_i''}(ENC_{K_i}(a_i))$ to HE
4: KH sends $K$ to HE
5: Choose keys $K_i''' \in [0, 2^{l-i}]$ {by HE}
6: If $K_i = 0$: $ENC_{Kf(i)}(a_i) := ENC_{K_i'''+K_i''}(ENC_{K_i}(a_i)) = ENC_{K_i''}(ENC_{K_i}(a_i)) + K_i'''$ $(mod\ 2^{l-i})$ {by HE}
7: If $K_i = 1$: $ENC_{Kf(i)}(a_i) := -ENC_{K_i''}(ENC_{K_i}(a_i)) + K_i'''$ $(mod\ 2^{l-i})$ {by HE}
8: HE sends $ENC_{Kf(i)}(a_i)$ to EVH and $K_i'''$ to KH
9: If $K_i = 0$: $Kf(i) := K_i'' + K_i'''$ $(mod\ 2^{l-i})$ {by KH}
10: If $K_i = 1$: $Kf(i) := -K_i''^{-1} - 1 + K_i'''$ $(mod\ 2^{l-i})$ {by KH}
11: $Kf := (\sum_{i \in [0, l-1]} 2^i \cdot Kf(i))$ $(mod\ 2^l)$ {by KH}
12: $ENC_{Kf}(a) := (\sum_{i \in [0, l-1]} 2^i \cdot ENC_{Kf(i)}(a))$ $(mod\ 2^l)$ {by EVH}
13: return $(ENC_{Kf}(a), Kf)$

---

$K_i'''$. The KH does not obtain any encrypted value at all. The HE obtains $ENC_{K_i''}(ENC_{K_i}(a_i))$ but not $K_i''$.

For correctness we show that $ef_i = a_i + kf_i \mod 2$ for a bit $i$ using a case distinction for bit $K_i$. We have by definition that $Kf(i) := K_i'' + K_i'''$ and $ENC_{K_i''}(ENC_{K_i}(a_i)) = (a_i \oplus K_i) + K_i''$ $(mod\ 2^{l-i})$.

Assume $K_i = 0$:
Then $ENC_{K_i}(a_i) = a_i \oplus 0 = a_i$
This implies $ENC_{K_i''}(ENC_{K_i}(a_i)) = ENC_{K_i''}(a_i)$. Therefore:

$$\begin{aligned}
& a_i + Kf(i) \ (mod\ 2^{l-i}) \\
=\ & a_i + K_i'' + K_i''' \ (mod\ 2^{l-i}) \ \text{(by Line 6 in XORToADD and using } ENC_{K_i}(a_i) = a_i \text{)} \\
=\ & (ENC_{K_i''}(a_i)) + K_i''' \ (mod\ 2^{l-i}) \ \text{(by definition of additive encryption)} \\
=\ & ENC_{K_i'''}(ENC_{K_i''}(ENC_{K_i}(a_i))) \ \text{(by definition of additive encryption)} \\
:=\ & ENC_{Kf(i)}(a_i)
\end{aligned}$$

Assume $K_i = 1$:
Then $ENC_{K_i}(a_i) = a_i \oplus 1 = 1 - a_i$ and $K_i''^{-1} := 2^{l-i} - K_i''$. We shall use that $-2^{l-i} + K_i''$ $(mod\ 2^{l-i})$ is congruent to $K_i''$ $(mod\ 2^{l-i})$, ie. $-2^{l-i} + K_i''$ $(mod\ 2^{l-i}) = K_i''$ $(mod\ 2^{l-i})$. Using this:

$$\begin{aligned}
& a_i + Kf(i) \ (mod\ 2^{l-i}) \\
=\ & a_i - K_i'' - 1 + K_i''' \ (mod\ 2^{l-i}) \ \text{(by Line 10 in XORToADD)} \\
=\ & -(1 - a_i + K_i'') + K_i''' \ (mod\ 2^{l-i}) \ \text{(rearranging)} \\
=\ & -(1 \oplus a_i + K_i'') + K_i''' \ (mod\ 2^{l-i}) \ \text{(Since } 1 - a_i = 1 \oplus a\text{)} \\
=\ & -(ENC_{K_i}(a_i) + K_i'') + K_i''' \ (mod\ 2^{l-i}) \\
=:\ & ENC_{Kf(i)}(a_i)
\end{aligned}$$

$\square$

## 2.2. Purely Additive ↔ Additive Modulo Encryption

Taking the modulo $2^l$ of the ciphertext and key yields the modulo encryption using a purely additive encryption, ie. the EVH computes $ENC_{K'}(a) = (a + K) \mod 2^l$ and the KH computes $K' = K \mod 2^l$ as shown in Algorithm AddToAddMod.

---
**Algorithm 3** AddToAddMod(encrypted value $ENC_K(a)$, key $K$)

1: return $(ENC_K(a) \mod 2^l, K \mod 2^l)$

---

For the opposite conversion we present two variants: The first one is very fast, but it puts constraints on the minimum key size. The second method is slower. It requires conversion to XOR encryption but comes with no restriction on the used keysize for encryption.

### 2.2.1. Fast Conversion

The key idea is to encrypt the additive modulo encrypted number with a key using less bits in an additive manner. This requires that the encrypted secret is (significantly) smaller than the used key size to ensure sufficient statistical security.

We convert an additively encryption $ENC_K(a)$ (using modulo) with $K$ of $l$ bits to an additive encryption without modulo $ENC_{K'}(a)$ with a key $K'$ of $l' = l-3$ bits. This allows simple conversion by subtracting a partially encrypted key $K$, ie. it leaks information about the highest order bits of $K$. Therefore, to ensure (statistical) security of $a$ one would choose $l$ significantly larger than the number of bits of $a$. We assume that $a$ has less than $l-3$ bits.

---
**Algorithm 4** AddModToAddFast(encrypted value $ENC_K(a)$, key $K$)

1: Choose key $K' \in [0, 2^{l'}]$ with $l' = l-3$ {by KH}
2: $ENC_{-K'}(K) := K - K'$ {by KH}
3: KH sends $ENC_{-K'}(K)$ to EVH
4: $ENC_{K'}(a) := (ENC_K(a) - ENC_{-K'}(K)) \mod 2^l$ { by EVH}
5: return $(ENC_{K'}(a), K')$

---

**Theorem 3.** *Algorithm 4 converts correctly and securely from additive modulo encryption to additive encryption.*

*Proof.* Security follows since none of the three parties exchange an encrypted value or a key that allows any of them to decrypt a value.
For correctness we consider two cases depending on the impact of the modulo for the $ENC_K(a)$ $(mod\, 2^l)$. We also assume that $a \geq 0$:
i) $a + K < 2^l$:
$ENC_K(a) - ENC_{-K'}(K) \ (mod\ 2^l) = a + K - K + K' \ (mod\ 2^l) = a + K'$.
Here we have that $a + K' < 2^l$ since the length of $K'$ is $l' = l-3$ bits and that length of $a$ is at most $l-3$.

ii) $a + K \geq 2^l$:
This implies that $ENC_K(a) = a + K - 2^l$. Therefore, $ENC_K(a) - ENC_{-K'}(K) = (a + K - 2^l - K + K') = (a + K' - 2^l) \ (mod \ 2^l) = a + K'$. The last equation is due to congruence in modular arithmetic. □

*2.2.2. Slow Conversion*

The key point is to take care of the carry bit that might have been chopped off by the modulo. This is done by prepending it (additively encrypted) to the additive modulo encryption. We compute the carry bit $le := c_l = LessZero(ENC_K(a), K)$. to In Algorithm AddModToAddSlow an additively encryption $ENC_K(a)$ (using modulo) is converted to an additive encryption without modulo $ENC_{K'}(a)$ with a key $K'$ of $b > l$ bits.

---
**Algorithm 5** AddModToAddSlow(encrypted value $ENC_K(a)$, key $K$)
---
1: $(ENC_{K'''}(le), K''') := LessZero(ENC_K(a), K)$ {Compute carry bit $c_l = le$}
2: $(ENC_{K''}(le), K'') := XORToAdd(ENC_{K'''}(le), K''')$ {Encrypt just additive, no modulo in $XORToAdd$}
3: $ENC_{Kf}(a) := ENC_{K''}(le) \cdot 2^l + ENC_K(a)$ {by EVH, prepend additive encryption of $le$}
4: $Kf := K'' \cdot 2^l + K$ {by KH, prepend key $K''$}
5: return $(ENC_{Kf}(a), Kf)$
---

**Theorem 4.** *Algorithm 5 converts correctly and securely from XOR encryption to additive encryption.*

*Proof.* Security follows since none of the three parties exchange an encrypted value or a key that allows any of them to decrypt a value.
Assume $a + K \geq 2^l$ implying the carry bit $c_l = 1$: Adding $ENC_K(a) + ENC_{K''}(le) \cdot 2^l = (a + K)(mod 2^l) + 2^l + K'' = a + K + K'' = ENC_{Kf}(a)$.
Assume $a + K < 2^l$ then $c_l = 0$: Adding $ENC_K(a) + ENC_{K''}(le) \cdot 2^l = (a + K)(mod 2^l) + K'' = a + K + K'' = ENC_{Kf}(a)$.
□

## 3. Logical Operations

We begin by stating an algorithm for the Hamming distance (Section 3.1). The Hamming distance is helpful for computing large fan-in gates (Section 3.2). The comparison of a number with zero (Section 3.3) is an application of large fan-in gates, whereas less than zero relies on using algorithm AddToXOR (Section 3.4). Comparing a number to zero is a necessary sub-procedure in our algorithm for computing carry bits given the sum of two numbers and one summand (Section 3.5).

### 3.1. Hamming Distance and 0-Test

For several logical operations we use the Hamming distance of an encrypted value and its key. The idea is that the Hamming distance is zero if and only if the confidential value is zero, ie. $a = 0 \Leftrightarrow ENC_K(a) = K$ (for all our encryptions schemes). In general, the Hamming distance gives the number of distinct bits for two numbers $x, y$ of $l$ bits. It is defined as $H(x, y) := \sum_{i \in [0, l-1]} x_i + y_i - 2x_i y_i$. We have that $a = 0 \Leftrightarrow H(ENC_K(a), K) = 0$, where $ENC_K(a)$ can be an arbitrary encryption scheme. The computation of $H(x, y)$ is straightforward, thus no pseudocode is given for algorithm $HAM(x, y)$ that computes $H(x, y)$, ie. it returns the additively encrypted result $(ENC_K(H(x, y), K)$. In the computation of $x_i \cdot y_i$ we treat the encrypted values $x_i = e_i$ and keys $y_i = k_i$ as secrets. Thus we must encrypt them beforehand. Algorithm $HAM(x, y)$ can be implemented in 3 rounds, ie. one for key dissemination, since we 'double' encrypt all encrypted values and also encrypt key bits and 2 rounds to conduct the multiplication $x_i \cdot y_i$. An alternative implementation requiring less communication but 5 rounds, replaces the multiplication with an $AND$ and uses conversation from XORtoADD. The number of bits (assuming preshared keys) are given by one $AND$ for each of the l bits, yielding a total of $5l$ and $2l$ bits to compute XORtoADD.

### 3.2. Large Fan-In Gates

We present three methods, one based on Hamming distances, one based on recursive computation using a procedure for fan-in gates of some fixed width and one that combines ideas of both. For Boolean Circuits, the AND of any number $w$ of terms, ie. $\wedge_{i \in [0, w-1]} a_i$ can be computed in $O(\log^* l)$ rounds using the Hamming Distance. The idea is that we sum up all $w$ negated bits $a_i$. The AND of all of them is true, if all $a_i$'s are one, ie. the sum of negated bits is zero. Since we want just a single bit, we have to repeatedly apply the Hamming Distance. Assume we start with encrypted value 1000 and key 10. Then $HAM(1000, 10)$ returns an encrypted value of 2 to the EVH and the key to KH, since the two numbers differ in two bits, ie. the second and fourth. Say, the EVH gets 11 and the KH 1. Computing $HAM(11, 1)$ yields just a single encrypted bit. This idea is realized in Algorithm FanInHamming. It is important to convert from additive to XOR encryption, since otherwise the number of bits that differ for the encryption and the key is not necessarily decreasing due to carry bits.

**Theorem 5.** *Algorithm 6 runs correctly and securely.*

*Proof.* Security follows from security of AddToXOR (Theorem 1), since none of the three parties exchange an encrypted value or a key that allows any of them to decrypt a value.
For correctness observe that the hamming distance of a number with $l$ bits is a number of $\log l$ bits. By definition of the $\log^*$ function (see Definition 2 in [25]). After at most $\log^* l$ iterations, we are left with just a single bit. □

**Algorithm 6** FanInHamming(encrypted values $ENC_{K_i}(a_i)$, keys $K_i$ with $i \in [0, w-1]$)

1: $E_\neg := \neg ENC_K(a)$ {by EVH, bitwise negation}
2: $(ENC_{K^0}(H_{(0)}), K^0) := AddToXOR(HAM(E_\neg, K), K')$ {Result of Hamming distance converted to XOR encryption}
3: **For** $i = 1$ **to** $\lceil \log^*(l_E + 1) \rceil$ **do**
4: $\quad (ENC_{K^i}(H_{(i)}), K^i) := AddToXOR(HAM(ENC_{K^{i-1}}(H_{(i-1)}), K^{i-1}), K'^{i-1})$ {number of bits of $K_i$ and encryption is reduced logarithmically}
5: **return** $(\neg ENC_{K^{\lceil \log^* l_E \rceil}}(H_{(\lceil \log^* l_E \rceil)}), K^{\lceil \log^* l_E \rceil})$ {EVH negates its value}

To compute ANDed terms with many variables, JOS allows to trade round complexity and message size. More precisely, fan-in gates of size *base* can be computed with messages of size $2^{base}$ in two rounds. Thus, we partition the expression of $l_E$ bits into $l_E/base$ terms of size *base*. This is followed by a recursive computation and the $AND$ of all partial results as shown in Algorithm FanInBase. For notational convenience assume $l_E$ is a power of *base*, ie. $l_E := base^j$ for some integer $j \geq 0$. The total number of rounds is then $2 \log_{base} l_E$.

**Algorithm 7** FanInBase(encrypted values $ENC_K(a)$, keys $K$)

1: **while** $l_E > 1$ **do**
2: $\quad \forall i \in [0, l_E/base - 1]$ in parallel: $(E'_i, K'_i) := AND(\forall_{j \in [i \cdot base, (i+1) \cdot base - 1]}(E_j, K_j))$
3: $\quad E := E'$ {by EVH, new bit length $l_E$ of $E$ is only a fraction $1/base$ of $E'$}
4: **end while**
5: **return** $E, K$

An even faster algorithm FanInBoth could be achieved when combining ideas from both algorithms. FanInHamming requires $5 \cdot \log^* n$ rounds with messages of size $l$ (with $l$ being the keysize), whereas FanInBase requires $2 \log_{base} l$ rounds with messages of size up to $base \cdot 2^{base}$. The exponential growth of the message size makes FanInBase only practical for small values of *base*. To combine both algorithms we first compute the Hamming Distance recursively for $i < \log^* n$ recursions, yielding a number of length $\log^{(i)} l$ bits (with $\log^{(i)} l$ being the $i$ times iterated logarithm) and then run algorithm FanInBase. The total runtime is given by $5 \cdot i + 2 \log_{base} \log^{(i)} l$. This allows to trade between communication and round complexity. In practice, the optimal choice of parameters $i, base$ depend on network parameters such as bandwidth and latency and the hardware of the machines involved.

*3.3. Equality to Zero*

Given confidential value $a$ XOR encrypted we compute whether $a$ equals zero, ie. $a \stackrel{?}{=} 0$, such that the EVH holds the encryption $ENC_{K_{eq}}(a \stackrel{?}{=} 0)$ and the KH the key $K_{eq}$.

---

**Algorithm 8** FanInBoth(encrypted values $ENC_K(a)$, keys $K$)

1: $E_\neg := \neg ENC_K(a)$ {by EVH, bitwise negation}
2: $(ENC_{K'}(H'), K') := HAM(E_\neg, K)$ {Hamming distance using negated bits}
3: $l_{EH}$ = number of bits to store Hamming distance, ie. $\lfloor (\log l_E) \rfloor + 1$
4: $(E'', K'') := AND(\forall_{i \in [0, l_{EH}-1]}(ENC_{K'}(H')_i, K'_i))$
5: return $(E'', K'')$

---

Protocol *EqualZeroFan* uses the fact that the secret $a$ is equal to 0, if all its bits $a_i$ equal 0. Thus, $a$ is zero, if the AND of all negated bits is one, ie. $a \stackrel{?}{=} 0 \iff \wedge_{i \in [0, w-1]}(\neg a_i)$.

---

**Algorithm 9** EqualZeroFan(encrypted value $ENC_K(a)$, key $K$)

1: $base := 6$ {arbitrary value}
2: $E_\neg := \neg ENC_K(a)$ {by EVH, bitwise negation}
3: return $FanInHamming(E_\neg, K)$ (or other algorithm, such as $FanInBoth$)

---

*3.4. Less Zero Comparison*

For a secret $a$ encrypted additively with key $K$, ie. $ENC_K(a) = a + K$ (potentially, $\mod 2^l$) we compute whether $a$ is less than zero, ie. $a \stackrel{?}{<} 0$. We assume that $a$ is given in two's complement. Our algorithm compares the key and the encrypted value and returns the (encrypted) key bit of the most significant bit where the encryption and the key differ. If there is no different bit ($a = 0$) then we return (encrypted) 0. Our (Monte Carlo) algorithm LessZero returns a result with a probability that can be made arbitrary large, if keys are chosen uniformly at random. Alternatively, one might avoid using keys which have higher order bits all equal to one. Then algorithm LessZero always returns the correct result, but it only ensures statistical security.

The computation of the comparison protocol *LessZero* proceeds recursively: If the first half of all bits differs from 0 then ignore the second half of bits and return the MSB of the first half. Otherwise ignore the first half and return the MSB of the second half.

We define the minimal number of bits to encode $|a|$ as $n_a := \lfloor \log(\max(|a|, 1)) \rfloor + 1$ with $n_a \in [1, l-1]$. Let $n_s := l - n_a \geq 1$ be the number of bits to represent the sign. In standard two's complement $n_s$ is at least one. However, by restricting the maximum allowed size of $|a|$ or increasing the number of bits $l$, $n_s$ can be made arbitrarily large. We assume $n_s > 1$.

**Theorem 6.** *For a single run of algorithm* LessZero *holds* $p_{err} \leq 1/2^{n_s}$.

*Proof.* We discuss three cases: $a > 0, a = 0$ and $a < 0$. We assume $a$ is represented in two's complement. If $a$ equals zero then the computation is correct, since it returns simply the result of algorithm EqualZero. Recall that

**Algorithm 10** LessZero(encrypted value $ENC_K(a)$ with $l > 1$ bits, key $K$)

---
1: $l :=$ number of bits of the encrypted value
2: **if** $l > 1$ **then**
3:     $El := ENC_K(a)/2^{l/2}; Kl := K/2^{l/2}$ {by EVH/KH, left half of all bits}
4:     $Er := ENC_K(a) \mod 2^{l/2}; Kr := K' \mod 2^{l/2}$ {by EVH/KH, right half of all bits, *locally*}
5:     $(ENC_{K'}(b), K') := EqualZero(El, Kl)$
6:     $neb := 1 - ENC_{K'}(b)$ {by EVH}
7:     $nk := 2^l - K'$ {by KH}
8:     {Compute: $b \cdot LessZero(El, Kl) + (1-b) \cdot LessZero(Er, Kr)$}
9:     $(ENC_{K^1}(lle0), K^1) := LessZero(El, Kl)$
10:    $(ENC_{K^2}(rle0), K^2) := LessZero(Er, Kr)$
11:    $(ENC_{K^3}(rble0), K^3) := MUL((ENC_{K'}(b), K'), (ENC_{K^1}(lle0), K^1))$
12:    $(ENC_{K^4}(rbre0), K^4) := MUL((neb, nk), (ENC_{K^1}(rle0), K^1))$
13:    **return** $(ENC_{K^3}(rble0) + ENC_{K^4}(rbre0), K^4), K^3 + K^4)$
14: **else**
15:    $ENC_{K'}(b), K' := EqualZero(ENC_K(a), K)$
16:    $ENC_{K'}(\neg b), K' := \neg ENC_{K'}(\neg b), K'$ {by EVH}
17:    $ENC_{K''}(bk), K'' := MUL((ENC_{K'}(b), K'), (0, K))$
18:    **return** $(ENC_{K''}(bk), K'')$
19: **end if**
---

the ciphertext is the addition of the key and the plaintext and that a carry bit $i$ is one if the sum of bit $i$ of a secret, the previous carry $i-1$ and the key bit $i$ is at least two. For a positive number $a > 0$ for the MSB the key bit is zero. This holds if there are no carry bits being one, since in this case all bits differing in the key and ciphertext are due to some bits of $a$ being one and the corresponding key being zero. Assume there is a carry bit being one. Since in two's complement all leading bits (bits in front of the $n_a$ bits of $a$) are zero, a carry bit due to some $a_i + K_i > 1$ (for $i < n_a$) propagates from bit to bit until it reaches a key bit $j > i$ with $K_j + a_j + c_j = 0$. Consider the propagation from the largest $i^* < n_a$ such that $a_{i^*} + K_{i^*} > 1$. It yields the bit $j^*$ with $K_{j^*} = 0$ and $a_{j^*} = 0$ being the MSB, if there exists a $j$ such that $K_j = 0$. If there is no such $j$ for an $i^* < n_a$, the sign bit is not computed correctly. Thus, it is sufficient if one of the key bits $K_j$ with $j \in [n_a, l-1]$ is zero. The probability that none of the key bits is zero is $1/2^{l-n_a} \leq 1/2^{n_s}$.

Consider a negative number $a < 0$ in two's complement. In case a key bit $i \geq n_a$ is one (or the carry bit $c_{n_a}$ is one), the propagation of the resulting carry bit $c_k$ with $k > n_a$ does not stop before $j^* = l$, since all bits $a_j$ are 1 for $j \geq i \geq n_a$. Thus, for $c_{n_a} = 0$ for the MSB at position $i \geq n_a$ holds $K_i + a_i = 2$, ie. the key bit is 1. For $c_{n_a} = 1$ we have that $a_{n_a-1} = 0$ since $a$ is negative and in two complement, this implies $K_{n_a-1} + a_{n_a-1} + c_{n_a-1} = K_{n_a-1} + c_{n_a-1} = 2$, ie. the key bit is 1. Thus, the computation is correct if either $c_{n_a} = 1$ or there exists a key bit $K_j$ equal to one for $j \in [n_a, l-1]$. The probability that none of the key bits is one is $1/2^{l-n_a} \leq 1/2^{n_s}$. □

When using a single sign bit, ie. $n_s = 1$, the error probability is one half according to Theorem 6. This is no better than guessing. This implies one must use more than one sign bit. The error probability can be made arbitrarily small storing more sign bits. Alternatively, (or additionally) we can execute Algorithm *LessZero* several times, ie. $n_e$ times, using different encryptions of the same plaintext. The final result for $a \stackrel{?}{<} 0$ is the value returned by the majority of executions of algorithm *LessZero* as shown in algorithm *MultiLessZero*. To compute the majority we also use procedure *LessZero* (but) using a large(r) number of sign bits.

---

**Algorithm 11** MultiLessZero(encrypted value $ENC_K(a)$, key $K$)

---
1: Choose $n_e$ keys $K^i$ uniformly at random {by KH}
2: KH sends $(K^{-1} + K^i)$ to EVH
3: $ENC_{K^i}(a) := ENC_K(a) + (K^{-1} + K^i)$ {by EVH}
4: In parallel: $(E^{i'}, K^{i'}) := LessZero(ENC_{K^i}(a), K^i)$
5: {Compute $n_e/2 - \sum LessZero(ENC_{K^{i'}}(a), K^{i'})$}
6: $Es := n_e/2 - \sum_{i \in [0, n_e-1]} E^{i'}$ {by EVH, with a key of length $1 + \lfloor \log n_e \rfloor + n_b$}
7: $Ks := -\sum_{i \in [0, n_E-1]} K^{i'}$ {by KH, with a key of length $1 + \lfloor \log n_e \rfloor + n_b$}
8: return $LessZero(Es, Ks)$

---

**Theorem 7.** *Algorithm* MultiLessZero *returns the correct result with probability at least* $1 - 2e^{-n_e(2^{n_s-1}-1)/6}$ *for* $n_s \geq 2$.

*Proof.* Using Theorem 6 it holds for the error probability for one execution of algorithm *LessZero*: $p_{err} \leq 1/2^{n_s}$. When doing $n_e$ executions we expect $\mu = n_e/2^{n_s}$ errors. Let $X \in \{0, n_e\}$ be the number of errors. Since we return the majority of results of algorithm *LessZero*, the computed result by *MultiLessZero* is correct as long as $X < n_e/2$. We use a Chernoff bound $p(X \geq (1+\rho)\mu) \leq e^{-\rho^2 \mu/3}$. We have $\mu \cdot (1 + \rho) = n_e/2$ for $\rho = n_e/(2 \cdot \mu) - 1 = 2^{n_s-1} - 1$. Thus, $p(X \geq (1+\rho)\mu) \leq e^{-(2^{n_s-1}-1)^2 \mu/3} \leq e^{-n_e(2^{n_s-1}-1)/6}$. For computing the majority using *LessZero* we use $n'_s := n_b$. Thus, the error probability for this execution of *LessZero* is given by $1/2^{n_b}$. We use $n_b = \lceil \log(e^{-n_e(2^{n_s-1}-1)/6}) \rceil$ bits to get an error bound of algorithm MultiLessZero of $2e^{-n_e(2^{n_s-1}-1)/6}$. □

*3.5. Carry Bits*

We compute the carry bits $c_i$ that stem from encryption, ie. addition of $a$ and $K$. Since $a$ is secret we can only use the sum, ie. the encryption, and the key to calculate the carry bits. We propose two methods to compute a carry bit: One is based on using a Boolean expression for each carry bit, which yields rather long expressions. The second technique uses comparisons, ie. less than zero.

*3.5.1. Boolean Expression based*

Carry bits can be computed using Boolean expressions, ie. $c_0 := (\neg k_0) \wedge e_0$, $c_i := (k_i \wedge e_i \wedge (\neg c_{i-1})) \vee (c_{i-1} \wedge (\neg k_i) \wedge (\neg e_i)) \vee (c_{i-1} \wedge k_i \wedge e_i)$. The recursive definition can be expanded by substitution into a single expression for each carry bit $c_i$ depending only on bits of the key and the encrypted value. All carry bits can then be computed in parallel. The length of the longest expression for a carry bit, ie. for $c_{l_E}$ is of order $l_E$.

*3.5.2. Comparison based*

The idea is to compare parts of the encrypted value and the key for each bit. We use (and prove) that $c_i = 1 \iff ENC_K(a) \mod 2^i < K \mod 2^i$ and $c_i = 0$ otherwise. In Algorithm *CarryBits* the modulo of the ciphertext and key can be computed without communication. However, for the comparison, ie. $ENC_K(a) \mod 2^i - K \mod 2^i < 0$, we must encrypt $K \mod 2^i$ using a key $K'_i$. The EVH computes $E'_i := (ENC_K(a) \mod 2^i(-K \mod 2^i + K'_i))$ mod $2^i$. Finally, we do the comparison of the modulo values yielding the carry bits.

---

**Algorithm 12** CarryBits(encrypted values $ENC_{K_i}(a_i)$, keys $K_i$)

1: Compute in parallel for all $i \in [0, l-1]$:
2:     $(ENC_{K''_i}(c_i), K''_i) := LessZero(ENC_K(a) \mod 2^i, K \mod 2^i)$
3: return for $i \in [0, l-1]$: $(ENC_{K''_i}(c_i), K''_i)$

---

**Theorem 8.** *For each carry bit $c_i$ Algorithm* CarryBits *returns the correct result with probability at least that of LessZero.*

*Proof.* We have $c_i = 1 \iff ENC_K(a) \mod 2^i < K \mod 2^i$ and $c_i = 0$ otherwise. To see this, assume that $c_i = 0$, ie. there is no carry over when adding the last $i$ bits of the key and the plaintext. Mathematically speaking, $ENC_K(a) \mod 2^i = (a \mod 2^i) + (K \mod 2^i)$. Since $a \mod 2^i \geq 0$ it follows that $ENC_K(a) \mod 2^i \geq K \mod 2^i$. Assume $c_i = 1$, then we must account for the carry bit by adding it, ie. $ENC_K(a) \mod 2^i + 2^i = (a \mod 2^i) + (K \mod 2^i)$. Since $2^i > a \mod 2^i$, we have $ENC_K(a) \mod 2^i < K \mod 2^i$.
Security and probability of correctness follow from protocol LessZero. □

## 4. Numerical Operations

Before diving into numerical operations we discuss a supporting algorithm, ie. in Section 4.1 we show how to obtain the index of the most significant bit(MSB) of a number. This algorithm is used for computing the division and logarithm of a confidential number using a Taylor series (Section 4.2). Section 4.3 states procedures for dividing and multiplying an encrypted value by a non-confidential value. Section 4.4 discusses trigonometric functions, ie. how to calculate the sine, cosine and tangent function.

*4.1. Index of MSB*

The index of the most significant bit (MSB) of a positive number $a$ encrypted additively (potentially, $\mod 2^l$) is calculated investigating bit by bit. We begin from the highest order bit and move towards the lowest order bit. For each bit $i$ of $a$ we check if it is set, ie. $a_i = 1$. This is done by subtracting $2^i$ from the plaintext for all $i$ (using two's complement) and checking if the result of the subtraction is less than zero. We assume that $a < 2^{l-3}$ so that we can conduct the subtraction and get a result in two's complement, ie. the top most bit(s) can be used as sign bits. The comparison yields a sequence of $l$ bits $le_{l-1}|le_{l-2}|...|le_{l-0}$ of the form $11...100...0$, such that all bits $le_i$ are 1 for $i$ with $2^i > a$ and 0 otherwise. We can compute the sum $s$ of all bits $le_i$. Thus, the index of the MSB is then $l-3-s$. Sometimes, we are interested in computing the power of the index $2^{l-3-s}$ which is computed by XORing one bit with the next, ie. $pos_i := le_i \oplus le_{i+1}$, yielding a number $00...0100...0$ through concatenation, ie. $2^{l-s} := pos_l|pos_{l-1}|...|pos_0$. Depending on the application, we might only compute the MSB and not $2^{MSB}$.

---

**Algorithm 13** IndexMSB(encrypted value $ENC_K(a)$, key $K$)

---
1: Compute in parallel for all $i \in [0, l-3]$:
2:    $(ele_i, kle_i) := LessZero(ENC_K(a - 2^i) \mod 2^l, Kl_i)$
3: $eMSB := l - 3 - \sum_{i \in [0,l-3]} ele_i$ {by EVH}
4: $kMSB := 2^{l-3} - (\sum_{i \in [0,l-3]} kle_i \mod 2^l)$ {by KH}
5: $epos_i := ele_i \oplus ele_{i+1}$ {by EVH, $i \in [0, l-3]$ and $ele_l := 0$}
6: $kpos_i := kle_i \oplus kle_{i+1}$ {by KH, $i \in [0, l-3]$ and $kle_l := 0$}
7: $ePow2MSB := epos_{l-2}|epos_{l-3}|...|epos_0$ {by EVH, concatenation of bits giving $2^{msb}$}
8: $kPow2MSB := kpos_{l-2}|kpos_{l-3}|...|kpos_0$ {by KH, concatenation of bits giving $2^{msb}$}
9: return $(eMSB, kMSB)$ and $(ePow2MSB, kPow2MSB))$ {Pow2MSB encrypted using XOR}

---

**Theorem 9.** *Algorithm* IndexMSB *returns the correct result with probability at least $p^l$, where $p$ is the success probability of LessZero for one call.*

*Proof.* We have that $a - 2^i < 0$ for all indexes $i$ that are larger than the index of the MSB $i^*$ and $a - 2^i \geq 0$ otherwise. Let the indicator $I_i = 1 \iff a - 2^i < 0$ and otherwise 0. Therefore we have that $l - 2 - \sum_{i \in [0,l-3]} I_i = l - 2 - \sum_{i \in [0,i^*-1]} 0 - \sum_{i \in [i^*,l-3]} 1 = l - 3 - (l - 3 - i^*) = i^*$. We also have that $I_i \oplus I_{i+1} = 1$ only for $i = i^* - 1$. Security and probability of correctness follow from protocol LessZero and the fact that we need a correct result for all $l$ bits. □

*4.2. Taylor Series: Division and Logarithm*

We present a technique to compute the Taylor series of any function $f$ on a secret value $a$. The Taylor series corresponds to a function that approximates function $f$. For a Taylor series we require a value $a_T$ for which we 'develop' the series. The error of the series depends on the properties of $f$ and typically the difference $a - a_T$ between the point used for developing the series and the point $a$ for which we wish to compute (an approximation of) $f$. Our technique is efficient as long as the Taylor series converges fast given that the secret $a$ and point for developing $a_T$ differ by less than 50%, ie. $|(a - a_T)/a_T| < 0.5$ and the function $f$ can be expressed using a scaled value $a' = a \cdot s$, ie. $f(a) = f(a' \cdot 1/s) = g(f(a'), f(1/s))$ for a function $g$ like addition or multiplication. For instance, for division we have $1/a = 1/a' \cdot s$ and, thus, $g$ is the multiplication function.

The technique relies on scaling the secret, which allows to use the same point for developing the series for any secret. More precisely, we scale the secret $a$ such that any scaled value lies within a fixed interval, where the lower and upper bound differs only by a factor of 2. We keep the scaling factor secret. This allows to choose the arithmetic mean (or any other value) out of this fixed interval as a point $a_T$ for developing any Taylor series (for any function $f$). Thus, this fixed value $a_T$ reveals nothing about the secret $a$ and does not have to be kept secret. In turn, this significantly simplifies computation, since large portions of the series can be computed on non-encrypted data, as we demonstrate in detail for division and logarithm.

For division we compute the inverse $1/a$ of a confidential value $a \in [1, 2^l - 1]$. This allows to compute any division of type $b/a$ with both values being confidential by simply multiplying $b$ and $1/a$.

Assume we are given a secret $a \in [1, 2^l - 1]$ and a scaled value $a' = a \cdot s$ of $a$ such that $a' \in [2^{l-1}, 2^l - 1]$ for some constant $l$. We develop the Taylor series around a fixed point $a_T$ lying in the middle of the interval, ie. $a_T := 2^{l-1} + 2^{l-2}$. The Taylor series for $1/a'$ around $a_T$ is given by $f(a') := 1/a_T - \frac{1}{2 \cdot a_T^2} \cdot (a' - a_T) + \frac{1}{6 \cdot a_T^3} \cdot (a' - a_T)^2 - \ldots = \sum_{i=0}^{\infty} d_i \cdot (a' - a_T)^i$ with constants $d_i := \frac{(-1)^n}{(i+1)! a_T^{i+1}}$. If we cut the series after $n_t$ terms, we can compute an upper bound on the error term $e_{n_t}$, using $|(a' - a_T)| \leq 2^{l-2}$:

$$\begin{aligned} |e_{n_t}| &\leq \max_{z \in [2^{l-1}, 2^l - 1]} \frac{1}{(n_t + 2)! z^{n_t+1}} \cdot (a' - a_T)^{n_t+1} \\ &\leq \frac{2^{(l-2) \cdot (n_t+1)}}{(n_t + 2)! \cdot 2^{(l-1) \cdot (n_t+1)}} \\ &= \frac{1}{(n_t + 2)! 2^{n_t+1}} \end{aligned}$$

To get 21-bit IEEE float precision we need $n_t = 7$. To get 52-bit double precision we need $n_t = 13$.

The Taylor series of $\log(a')$ around $a_T$ is given by $f(a') := \log(a_T) + \frac{1}{a_T} \cdot (a' - a_T) + \frac{1}{2 \cdot a_T^2} \cdot (a' - a_T)^2 + \ldots = \log(a_T) + \sum_{i=1}^{\infty} d_{i-1} \cdot (a' - a_T)^i$.

Let us discuss the computation of the Taylor series in more detail. We start by scaling the confidential value $a$ with value $s = l - 1 - \lfloor \log(a) \rfloor$ such that $a' = a \cdot s \in [2^{l-1}, 2^l - 1]$ always has $l$ bits. The scaling factor $s$ is computed in three steps starting from getting the most significant bit $Pow2msb$, which is encrypted using XOR of the key. For instance, for a secret 1011 with $l = 6$, we obtain (encrypted) $Pow2msb = 01000$. The second step reverses this bit pattern to get $RevPow = 00010$ by simple reordering of the bits. Finally, we change the encryption of $RevPow$ from XOR to additive. A key step is to compute the powers of $(a' - a_T)^i$ for $i \in [2, n_T]$. The first power is simply obtained by computing $a' - a_T$. The other powers can be computed using $\log l$ multiplications, ie. we iteratively square the value $t := (a' - a_T)$ to get $t^2$, $t^4$, $t^8$ and multiply the newly obtained squared term with all prior multiplication results. For instance, in the first round of multiplication we get $t^2$, in the second $t^2 \cdot t = t^3$, $(t^2)^2 = t^4$, in the third $t^4 \cdot t = t^5$, $t^4 \cdot t^2 = t^6$, $t^4 \cdot t^3 = t^7$ and $(t^4)^2 = t^8$. We might also use larger Fan-in Gates to compute all terms up to $t^{n_t}$ using a constant number of rounds.

*4.3. Multiplication and Division by a Non-confidential Value*

The product of confidential value $a$ by the non-confidential value $c$, ie. $a \cdot c$, can be computed without communication, if the EVH multiplies the encrypted value of $a$ and the KH computes the product of the key and $c$.
Division $a/c$ for two values $a, c$, where $c$ is non-confidential and $a$ is additively encrypted (without modulo) can be done in two rounds.
For $ENC_K(a)/c$ and $K/c$, it might not hold that $DEC_{K/c}(ENC_K(a)/c) = \lfloor (a+K)/c \rfloor - \lfloor K/c \rfloor$. The computation is off by one if and only if $rem(K, c) + rem(a, c) \geq c$.
We obtain the division however with arbitrary precision using scaling as follows. The EVH calculates $\lfloor (2^k \cdot ENC_K(a))/c \rfloor$ for an integer $k > \log c$. The KH computes $K' = \lfloor (2^k \cdot K)/c \rfloor$. It also chooses a random $K''$ and it computes $K'' - K'$ and transmits this to the EVH. In turn, the EVH calculates $ENC_{K''}(2^k \cdot a/c) = \lfloor 2^k \cdot ENC_K(a)/c \rfloor + (K'' - K')$. Finally, it shifts $ENC_{K''}(2^k \cdot a/c)$ by $k$ digits to get $ENC_{K''/2^k}(a/c)$. This can be done by discarding the last $k$ bits. The KH computes the final key $Kf = K''/2^k$ by discarding the last $k$ bits of $K''$.

**Theorem 10.** *For $2^{k-1} \geq c$ the computation is correct.*

Remark: At the expense of a longer proof one can also show that $2^k \geq c$ suffices.

*Proof.* Without scaling, the result is wrong if $rem(K, c) + rem(a, c) \geq c$. With scaling the result is wrong if and only if $(rem(2^k K, c) + rem(2^k a, c))/2^k \geq 1$. Since $2^{k-1} > c$ and $rem(x, c) < c$ (for any value $s$) the sum of remainders $rem(2^k K, c) + rem(2^k a, c)$ is at most $(c-1) + (c-1)$. Thus, the result is wrong if $2(c-1)/2^k \geq 1 \iff c - 1 \geq 2^{k-1}$. However, by assumption $2^{k-1} \geq c$ and thus, the result is correct. □

**Algorithm 14** DivisionAndLog(encrypted value $ENC_K(a)$, key $K$)

1: Assume $|a| < 2^{l/2}$
2: $s := \{$scaling factor of $a$ for fixed point number, eg. for 00101.010 it is $2^3\}$
3: $sc_0 := 1$, $sc_i := 2^{l/2 \cdot (i-1)}$ {scaling vector for numerical purposes}
4: $a_T := 2^{l/2-1} + 2^{l/2-2}$, $d_i := \frac{sc_i}{(i+1)! a_T^{i+1}}$, $n_t := 7$ {Terms in Taylor series}
5: $(emsb, kmsb), (ePow2msb, kPow2msb) := IndexMSB(ENC_K(a)/, K)$ {encrypted with XOR, should be within $[0, l/2 - 1]$}
6: $i \in [0, l/2 - 1] : eRevPow_i := ePow2msb_{l/2-i}$; $kRevPow_i := kPow2msb_{l/2-i}$ {by EVH/KH, Reverse bit order of Pow2MSB}
7: $i \in [l/2, l - 1] : eRevPow_i := 0$; $kRevPow_i := 0$ {by EVH/KH, discard $l/2$ bits}
8: $(ENC_{K'}(RevPow), K') := XORToAdd(eRevPow, kRevPow)$ {Additive encryption from XOR, $K'$ has $l$ bits}
9: $(ENC_{K''}(a') := MUL((ENC_K(a), K), (ENC_{K'}(RevPow), K'))$ {Scaling of $a$ $K''$ has $l$ bits}
10: $et(1) := ENC_{K''}(a') - a_T$ {by EVH}
11: $kt(1) := K''$ {by KH}
12: $(et(i), kt(i)) := SCALEDPOW((et(1), kt(1)), i, 2^{l/2})$ {scaled powers}
13: $eInv(a') := \sum_{i=0}^{n_t} d_i \cdot et(i)$ {by EVH}
14: $kInv(a') := \sum_{i=0}^{n_t} d_i \cdot kt(i)$ {by KH}
15: {Division: Multiply $1/a' = 1/a \cdot RevPow$ with $2^{l/2}/RevPow$ s.t. get $2^{l/2}/a$}
16: $(ENC_{K'''}(Pow), K''') := XORToAdd(ePow2msb, kPow2msb)$ {Additive encryption from XOR}
17: $(ENC_{K^4}(2^{l/2}/a) := MUL((eInv(a'), kInv(a')), (ENC_{K'''}(Pow), K'''))$ {Scaling of $a$}
18: $eInv(a) := ENC_{K''}(2^{l/2}/a)/2^{l/2}$ {by EVH}
19: $kInv(a) := K^4/2^{l/2}$ {by KH}
20: {Log}
21: $elog(a') := \log(a_T) + \sum_{i=0}^{n_t} d_{i-1} \cdot et(i)$ {by EVH}
22: $klog(a') := \log(a_T) + \sum_{i=0}^{n_t} d_{i-1} \cdot kt(i)$ {by KH}
23: {To get $\log(a)$ subtract $\log(RevPow)$ from $\log(a') = \log(a) + \log(RevPow)$}
24: $elog(a) := elog(a') - emsb$ {by EVH}
25: $klog(a) := klog(a') - kmsb$ {by KH}
26: return $(eInv(a)), kInv(a))$ and $(elog(a), klog(a)))$

**Algorithm 15** NonConfDivisionAdditiveEnc(encrypted value $ENC_K(a)$, key $K$, constant $c$)

1: {Constant $c$ is a fixed point number with scaling factor $s = 2^k$ for some $k$}
2: $divs := k + s$ {Scaling for division, $2^s$ is precision (scaling factor) of $a$}
3: $ENC_K(a/c) := (ENC_K(a) \cdot 2^{divs})/c \cdot 2^{-k}$ {by EVH, $\cdot 2^{-k}$ means shifting to the right if $k > 0$, otherwise to the left}
4: $K' := (K \cdot 2^{divs})/c \cdot 2^{-k}$ {by KH}
5: return $(ENC_K(a/c), K')$ and $(elog(a), klog(a)))$

*4.4. Trigonometric functions*

We compute three trigonometric functions, namely sine, cosine and tangent, given an additive encrypted value $a$, ie. $ENC_K(a) := a + K$ (without modulo). We discuss two methods. The second method illustrated for the tangent has the disadvantage that it needs special care to avoid divisions by 0 (or $\infty$). It illustrates the general principle of using a set of linear equations to evaluate complex function of a confidential value.

$$\sin(a) + \sin(K) = 2 \cdot \sin(\frac{a+K}{2}) \cos(\frac{a-K}{2})$$
$$\cos(a) + \cos(K) = 2 \cdot \cos(\frac{a+K}{2}) \cos(\frac{a-K}{2})$$

We describe only sin (cos is analogous). The KH can compute $\sin(K)$. The EVH computes $t_0 := 2 \cdot \sin(\frac{a+K}{2})$. To get the remaining term, the KH computes $K'' := -2K + K'$, where $K'$ is a randomly chosen key. It sends $K''$ to the EVH and $K'$ to the helper. The EVH computes $ENC_{-K+K'}(a) = (a + K) + K'' = a - K + K'$ and sends this to the helper. The helper computes $a - K$ and $t_1 := \cos(\frac{a-K}{2})$. Then we multiply (securely) $t_0$ and $t_1$ and subtract $\sin(K)$. Note, that the above formulas do not require integers. Generally, secret $a$ will not be an integer, but rather a fixed point number (and so will be the key and encrypted value). Algorithm 16 describes an approach that used fixed point integers.

For the second method, we use two encryptions of $a$ with keys $K$ and $K'$. We only discuss the tangent function. Note, that one could also compute the tangent using $\tan(x) = \sin(x)/\cos(x)$:

$$\tan(a + K) = \frac{\tan(a) + \tan(K)}{1 - \tan(a) \cdot \tan(K)} \iff$$
$$\tan(a + K) \cdot (1 - \tan(a) \cdot \tan(K)))$$
$$= \tan(a) + \tan(K) \qquad (2)$$
$$\tan(a + K') \cdot (1 - \tan(a) \cdot \tan(K')))$$
$$= \tan(a) + \tan(K') \qquad (3)$$

**Algorithm 16** Sine(encrypted values $ENC_{K_i}(a_i)$, keys $K_i$ with $i \in [0, w-1]$)

1: $c := 8; s := 2^c$ {$c$ is number of bits after the comma for a fixed point integer}
2: $sk := s \cdot \sin(K/s)$ {by KH}
3: Choose $K'$ randomly {by KH}
4: $K'' := -2 \cdot K + K'$ {by KH}
5: KH sends $K''$ to EVH and $K'$ to HE
6: $t_0 := s \cdot 2 \cdot \sin(\frac{a+K}{2 \cdot s})$ {by EVH}
7: $ENC_{-K+K'}(a) = (a + K) + K'' = a - K + K'$ {by EVH}
8: EVH sends $ENC_{-K+K'}(a)$ to HE
9: $ENC_{-K}(a) := ENC_{-K+K'}(a) - K' = a - K$ {by HE}
10: $t_1 := s \cdot \cos(\frac{a-K}{2 \cdot s})$ {by HE}
11: Choose $K''$ randomly {by HE}
12: $ENC_{K''}(t_1) := (t_1 + K'') \mod 2^{l+c}$ {by HE}
13: HE send $ENC_{K''}(s \cdot t_1)$ to EVH and $K''$ to KH
14: Choose $K''$ randomly {by EVH}
15: $ENC_{K'''}(t_0) := (t_0 + K''') \mod 2^{l+c}$ {by EVH}
16: EVH send $K'''$ to KH
17: $(eak, kak) := MUL((ENC_{K'''}(t_0), K'''), (ENC_{K''}(t_1), K''))$
18: $ekf := eak/s$ {by EVH}
19: $kf := kak/s + \sin(K) \cdot s$ {by KH}
20: return $(ekf, kf)$

Subtracting Equation (3) from (2) gives:

$$\tan(a + K) \cdot (1 - \tan(a) \cdot \tan(K)))$$
$$- \tan(a + K') \cdot (1 - \tan(a) \cdot \tan(K')))$$
$$= \tan(K) - \tan(K')$$
$$\tan(a) \cdot (\tan(a + K') \tan(K') - \tan(a + K) \tan(K))$$
$$= \tan(K) - \tan(K') + \tan(a + K') - \tan(a + K)$$
$$\tan(a) =$$
$$\frac{\tan(K) - \tan(K') + \tan(a + K') - \tan(a + K)}{\tan(a + K') \tan(K') - \tan(a + K) \tan(K)} \quad (4)$$

In Equation (4) all terms $\tan(\cdot)$ of the denominator $de_{tan} := \tan(K) - \tan(K') + \tan(a + K') - \tan(a + K)$ can be computed by the KH and EVH locally and then aggregated securely. The enumerator $no_{tan} := \tan(a + K') \tan(K') - \tan(a + K) \tan(K)$ can also be computed securely. For performance reasons, the division $de_{tan}/no_{tan}$ could be carried out using the helper and revealing $no_{tan}$ to it, though this might pose security risks: The helper could compute $1/no_{tan}$ using the decrypted value $no_{tan}$ and then we can compute the product $de_{tan} \cdot 1/no_{tan}$ on encrypted values.

To avoid division by zero (or by values close to zero) or division by infinity (or close to infinity[2]) we require the following conditions for some constants $k, c_2$ and $c_3$:

$$\begin{aligned}
|\tan(K)| \geq k \cdot c_2 \wedge |\tan(K')| &\geq k \cdot c_2 \\
|\tan(K) - \tan(K')| &\geq k \cdot c_2 \\
|\tan(a + K')| > 1/k \wedge |\tan(a + K)| &> 1/k \\
|\tan(K)| < c_3 \wedge |\tan(K')| &< c_3 \\
|\tan(a + K)| < c_3 \wedge |\tan(a + K')| &< c_3
\end{aligned} \quad (5)$$

The first inequalities (including inequality (5)) avoid division by small values. The last two avoid dealing with very large values in the nominator or denominator. Let us discuss the impact of these inequalities on the range of suitable keys. For the derivative of $\tan(x)$ holds $d\tan(x)/dx = 1/\cos(x)^2 \geq 1$. Thus, $|\tan(x)| \geq |x|$. Since $\tan(x)$ is periodic with period $\pi$, let us focus on the range $x \in [-\pi/2, \pi/2]$. This yields for the first two inequalities that each eliminates all keys $K$ with $|K| > k \cdot c_2$. The third means also that we eliminate another range of width $k \cdot c_2$. Continuing in this manner and assuming that all ranges for keys that must be excluded are disjoint, the range of non-useful keys becomes $3k \cdot c_2 + 2/k + 4 \cdot c_3$ out of a range $[-\pi/2, \pi/2]$ of width $\pi$ of possible keys. Setting $k = 1/\sqrt{c_2}$ and $c_2 = 0.0001$ and $c_3 < 1e19$ the probability to choose two suitable keys $K, K'$, when choosing them randomly is still 0.999. When using $n_p$ pairs of keys $K, K'$ the probability that all key pairs are non-suitable becomes $0.001^{n_p}$. To ensure that the above conditions on the tangent of keys and ciphertext are fulfilled the KH could repeatedly select keys and the EVH could tell the KH to reencrypt values until all conditions are fulfilled. However, biasing the choice of keys might lead to an insecure scheme.

Thus, our approach is to compute the tangent multiple times using different key pairs, but substitute dummy values for $\tan(\cdot)$ in case a party detects a violation upon the conditions. We select the results of all computations of $\tan(a)$ that did not violate any condition and return their average. More precisely, we let the KH choose a fixed number $n_p$ of pairs of keys $P_i = (K^i, K'^i)$. The KH sets the bit $S_K(i)$ equal 1 for all pairs $P_i$ for which the conditions on the keys are satisfied. Analogously, the EVH checks the conditions for $\tan$ on the encrypted values and sets bits $S_E(i)$ analogously to $S_K(i)$. Let $de_{tan}(i)$ and $no_{tan}(i)$ be the denominator and nominator in Equation (4) using the $i^{th}$ pair. For every $i$ with $S_K(i) = 1$ the KH uses 1 rather than $\tan(K^i)$ and 2 rather than $\tan(K'^i)$. For every $i$ with $S_E(i) = 1$ the EVH uses 3 rather than $\tan(a + K^i)$ and 7 rather than $\tan(a + K'^i)$. This yields a sequence $t_i := no_{tan}(i)/de_{tan}(i)$, where some $t_i$ might correspond to $\tan(a)$ and others might not do so due to a violation of the constraints (5). We compute the sum of all results weighted by the product of $1 - S_K(i)$ and $1 - S_E(i)$, which essentially sets invalid results $t_i$ to 0 and take the average of the weighted terms. We compute $(\sum_i no_{tan}(i)/de_{tan}(i) \cdot (1 - S_E(i)) \cdot (1 - S_K(i)))/(\sum_i (1 - S_E(i)) \cdot (1 - S_K(i)))$.

---

[2]Note, $\tan(x) = \infty$ for $x = (i + 1/2)\pi$ and integer $i$

## 5. Empirical Evaluation

We evaluated the overhead for equality and the sine function of 32bit unsigned integers compared to the state of the art in C++. We ran our experiments on an Intel Core i5. We compare in terms of computation and communication. For the result of sine we used a fixed point number with 24 bits precision.[3] Note, that to obtain a result of this precision requires higher precision for the computation on encrypted values. For a key $K$ having $b$ bits, a secret $a$ having $l$ bits we must compute (the sine) using $b+1$ rather than $l$ bits. To compute the sine function, we used a key of size 50 bits. We used 53 bits precision for the sine of the keys and encrypted value.[4]

| Operation | NoEncyrption | [17] with GRR(online) | EqualZeroFan(This Paper) |
|---|---|---|---|
| Equal to 0 | 0.0046[ys/op] | 254.44[ys/op/party] | 0.8955[ys/op/party] |
| Sine | 0.0515 | - | 0.0718 |

Table 1: Comparison of local computation using slowest party.

| | [17] with GRR(total) | EqualZeroFan (This Paper) |
|---|---|---|
| Equal to 0 | $\geq 10000$ ($\geq 6$) [total bits (rounds)] | ¡500 (25)[total bits (rounds)] |
| | Series expansion[14] | Sine (This Paper) |
| Sine | $> 4500$ (5) [total bits (rounds)] | $\leq$ <400 (5) [total bits (rounds)] |

Table 2: Comparison of total communication of all parties together (in bits) and rounds.

For equality we implemented the algorithm EqualZeroFan and the algorithm in [17] using Gennaro-Rabin-Rabin (GRR) [13, 19] relying on Shamir's Secret Sharing for secure additions/multiplications. [17] relies on an expensive pre-processing phase for each comparison. We implemented GRR with improvements [19] using the NTL library[5]. For GRR we used a 33 bit prime.

Computational performance is shown in Table 1, where we computed the average time per operation per party in microseconds using 10 million operations. We used the slowest party for each scheme (for GRR all parties perform the same computations). For computing equality to 0, we outperform GRR by about a factor of 300, while lagging behind non-encrypted computation by about a factor of 200. For computing the sine, we only lag behind non-encrypted computation by a factor of 1.4.

The amount of communication is given in Table 2. For comparison to zero we rely on pre-shared keys between each pair of parties, eg. by exchanging a single key and then using this key as input for a pseudo random number generator for all of the 10 Mio. operations. For [17] using GRR each party sends two messages with 66 bits each per multiplication for the online phase. Thus, for three parties one multiplication needs $6 \cdot 66 > 400$ bits. The preprocessing phase per operation of [17] (See Section 3.1) is very expensive. More precisely we need a (shared) random value $R$, its inverse $R^{-1}$ and other powers $R^i$ with $i \leq l = 32$, ie. the number of bits of the compared number. This computation requires at least $l = 32$ multiplications yielding a total of more than 10000 bits. Only some of these multiplications can be computed in parallel, eg. we need to compute $x^2, x^4, x^8, x^{16}$ and $x^{32}$ sequentially, requiring 5 multiplications. In the online phase we need another multiplication. In contrast, JOS needs 160 bits per multiplication in total. Our algorithm EqualZeroFan calls algorithm $FanInHam$ $1 + \log^* 32 = 5$ times. The first call needs $7l$ bits, ie. due to the Hamming distance yielding less than 250 bits. The others need $7 \log l$, $7 \log \log l$ etc. bits giving a total of less than 500 bits. One call to $FanInHam$ needs at most 5 rounds, yielding a total of 25 rounds.

---

[3]This corresponds to IEEE single-precision floats.
[4]This corresponds to IEEE double-precision floating point.
[5]http://www.shoup.net/ntl/

For sine we focused on fixed-point integer computation. We compared against [7] to compute individual terms of the standard series expansion. The series is given by $\sin(x) = \sum_{k=0}^{9} (-1)^k \cdot x^{2k+1}/(2k+1)!$ which still yields smaller precision than our scheme using IEEE 64-bit double values with 52 bits precision. We assumed that $x^{19}$ is computed in five consecutive multiplications, ie. i) $x^2 = x \cdot x$, ii) $x^4 = x^2 \cdot x^2$ (and $x^3$), iii) $x^8$, iv) $x^{16}$ and v) $x^{19} = x^{16} \cdot x^3$. An additional round is needed to compute the product of each term. The number of bits for a single multiplication are at least 300 for any known scheme existing prior to JOS, see evaluation in [23]. Thus, to compute 10 terms and the powers requires at least $15 \cdot 300$ bits and 5 rounds. In contrast, we only need to compute one multiplication and a few scaling operations and adjustments of keys (See Algorithm Sine). Overall this amounts to less than 400 bits.

## 6. Related Work

Recently, many systems have been made for general secure multi-party computation, eg. [11, 21, 16, 2] based on various schemes such as [28, 3, 13, 5]. We build upon a recent efficient scheme (JOS) [23] using at least three parties to enable complex numerical (and logical) operations. It is discussed in depth in Section 1.3. There is a significant body of literature using classical works such as the GRR scheme [13] using Shamir's Secret sharing or the BGW scheme [4] for (complex) numerical computations, eg. [6, 1]. Other work [9, 20, 17, 1] is based on a library of primitive operations such as addition and multiplication (and, thus, potentially also GRR or BGW) to construct more complex operations. This kind of 'BlackBox' model has also been captured and formalized in [10]. Homomorphic encryption has also been employed for multi-party computations focusing on numerical or logical operations, eg. [12, 27]. More specifically, [10, 12] adopt threshold homomorphic encryption.

Many schemes for logical and arithmetic functions, eg. [17, 9, 20] require protocols to compute (fast) unbounded fan-in AND gates. It is possible in (expected) constant number of rounds for arithmetic gates in [3] using Shamir's Secret Sharing. A number $a_i$ held by party $i$ as $ENC(a_i) = R_i \cdot a_i \cdot R_{i-1}^{-1}$ with $R_i$ being random matrices. We do not use ideas from [3]. Our approach to compute large fan-in AND gates is computationally more efficient than [3], but might require more communication.

Operations on fixed-point numbers such as scaling, comparison, multiplication and division are discussed in [7]. For division, [7] requires $2\Theta + 17 + 3\log(l)$ bits for a number of $l$ bits and a parameter $\Theta$ giving the number of iterations, ie. the precision of the result. For a floating point number with 23 bit precision, ie. an IEEE 32 bit floating point number, this results in at least 38 rounds and for 52 bits at least 44 rounds. For division, [7] builds upon the Goldschmidt approximation. In contrast, we apply a Taylor series. Our complexity is $O(\log(l \cdot n_T))$, where $n_T$ is the number of terms used in the Taylor series. For a 23 bit numbers we require at most 20 rounds, for double precision with 52 bits at most 30 rounds.

The first work to compute equality, comparison and discrete exponentiations in constant rounds was [9]. It relies on bit decompositions using $O(l \log l)$ multiplications, ie. obtaining a separate encrypted value for each bit of a secret as well as [3] to compute unbounded fan-in multiplication gates. Follow up work [20] improves on [9] by using bitwise shared secrets. The line of work [20, 9] is further enhanced by [17, 26]. Two (named) non-colluding parties and Paillier encryption are used in [26] for some operations. The state-of-the-art is [17] which leverages ideas from [26]. For equality [17] computes the Hamming Distance and compares it to zero using an $l$ degree polynomial with $l$ being the number of bits of a number. For "greater than" the idea is to check if the $l/2$ most significant bits are equal, and only if they differ check the $l/2$ least significant bits. This idea has been shown to work well. It has been used also in [12, 20]. The two party case for integer comparison is elaborated in [12]. Our work does not contain polynomials but one of our protocols also uses the idea of computing the Hamming Distance for large fan-in gates, which helps in computing equality to zero. One difference is that we change encryptions (between XOR and additive).

Floating point operations such as addition, multiplication, division, square root, logarithm and exponentiation are given in [1]. The logarithm is computed using the relation $\log(a) = 2\log e \cdot \arctan(\frac{a-1}{a+1})$ and a Taylor series approximation of arctan. The division is carried out using [7]. In contrast we compute the division using a Taylor series of $1/a$ using a fixed approximation point for all values $a$ and scaling of the divisor $a$. Though [1] argues against using a Taylor series (directly on log), we show that for the JOS scheme in combination with our technique this does not seem to hold. More precisely, we fix the evaluation point of the series. In fact, our complexity for logarithm equals the complexity for division. Since we use less rounds than [7] for division and [1] uses [7] for division in the computation of the logarithm, we are also faster than [1] (by more than 10 rounds). The square root [1] is computed iteratively using the Babylonian formula for $O(\log l)$ rounds. Division of floating point numbers is also based on [7]. Exponentiation is performed by doing a bitwise decomposition. Most ideas from [1] would be applicable for JOS. Very recently [14] most likely parallel to our work[6] published a work on integer and floating-point arithmetic. They discuss sine using a series approximation, power computation (in logarithmic number of rounds in the power) as well as logarithm using a fixed point polynomial. The usage of a fixed point polynomial is similar to us, since computation is also done with values in $[0, 1]$. [14] does not provide a bound on the precision of the approximation (as we do) but (also) provides a great level of detail to understand the protocol. Other work as focused on particular problems, eg. computation of the area of a triangle [18], graphs [15].

Some work compared quantitatively secure multi-party computation with respect to the function being a boolean circuit and the function being an arithmetic circuit over a large field [8]. This is also interesting in our context, since we show how to convert between bitwise XOR encryption (for Boolean circuits) and arithmetic encryption in a ring for the JOS scheme. Thus, one could use the JOS scheme with other schemes (compared in [8]) and pick the best algorithm for an operation. It has also been investigated whether one can disclose encrypted variables based on their values to speed up computation [24]. Such a scheme benefits from fast comparisons as well as fast operations involving only one encrypted term, which we provide in this work.

Division using homomorphic encryption (Paillier) is done in [27] as well as additive blinding assuming that the divisor is public. [27] performs approximate divisions, comparisons and min/max computations. In particular, they also use the fact that for a public divisor $d$ we have $(a + K)/d$ equals either $a/d + k/d$ or $a/d + k/d + 1$. However, [27] (and also prior work [6]) uses a comparison protocol to determine which of the two cases holds, which we do not require. In turn, we use scaling, ie. multiplication of a public number. This needs less rounds (and local computation) than comparison. The work of [6] also investigated upon primitives for MPC for integers with focus on comparison and truncation.

## 7. Conclusions

This work has given new approaches to compute complex functions with little computational overhead. Additionally, we have also presented several efficient protocols for logical operations. We believe that this work together with the JOS scheme paves the road to make secure multi-party computation practical in a number of (industrial) applications.

---

[6]The preprint of this manuscript [22] appeared before [14].